\documentclass[12pt,preprint]{aastex}

\begin{document}

\title {X-RAY AND INFRARED OBSERVATIONS OF TWO EXTERNALLY-POLLUTED WHITE DWARFS}

\author{M. Jura\altaffilmark{1}, M. P. Muno\altaffilmark{2}, J. Farihi\altaffilmark{3}, and B. Zuckerman\altaffilmark{1}} 

\altaffiltext{1}{Department of Physics and Astronomy and Center for 
Astrobiology, University of California, Los Angeles CA 90095-1562; jura, ben@astro.ucla.edu}
\altaffiltext{2}{Space Radiation Laboratory, California Institute of 
Technology, Pasadena CA 90025; mmuno@srl.caltech.edu}
\altaffiltext{3}{Department of Physics and Astronomy, University of Leicester, University Road, Leicester LE1 7RH, UK; jf123@star.le.ac.uk}

\begin{abstract}
With  {\it XMM-Newton} and the {\it Spitzer Space Telescope}, we obtain upper bounds to the X-ray fluxes from   G29-38
and GD 362, and the 70 ${\mu}$m flux from G29-38.  These data provide indirect evidence that G29-38 is accreting from a tidally disrupted asteroid: it is
neither accreting large amounts of hydrogen and helium nor is  its surrounding dusty disk  being replenished from a reservoir of cold grains experiencing Poynting-Robertson drag.    The upper bound to the X-ray flux from GD 362 is consistent with
the estimated rate of mass accretion required to explain its pollution by elements heavier than helium.   GD 362 also possesses 0.01 M$_{\oplus}$ of hydrogen, an anomalously large amount for a white dwarf with a helium-dominated atmosphere.   One possibility is that before the current disk was formed, this hydrogen was  accreted  from either ${\sim}$100  Ceres-like  asteroids or one large object.    An alternative  scenario which simultaneously explains all of GD 362's distinctive properties is that we are witnessing the consequences of the tidal-destruction of a single parent body that had internal water and was at least as massive as Callisto and probably as massive as Mars.  
\end{abstract}

\section{INTRODUCTION}

At least 1\% of  single white dwarfs with cooling ages less than 0.5 Gyr are known  to display both infrared excess emission and  photospheric elements heavier than helium (Farihi et al. 2009).  Since primordial heavy atoms gravitationally settle below the atmospheres in these stars, 
the observed heavies likely have been accreted  from  their circumstellar disks (Koester 2009).  Evidence is strong  that these disks are derived from the tidal-disruption of an asteroid or minor planet (Jura 2008).   If so, then the abundances in the atmospheres of white dwarfs can serve
as an indirect but uniquely powerful tool to measure the bulk composition of extrasolar planetary matter (Zuckerman et al. 2007).  For example, from the limited data now available, it appears that extrasolar asteroids resemble the inner Solar System with values of $n$(C)/$n$(Si) or $n$(C)/$n$(Fe) at least a factor of 10 below Solar  (Jura 2006, Farihi et al. 2009).    

Models of the infrared excess (Jura et al. 2007a) and measurements of  double-peaked calcium triplet emission (Gaensicke et al. 2006, 2007, 2008) both show that white dwarf disks have orbital radii of ${\sim}$10$^{10}$ cm even though these stars previously had radii of 
${\sim}$10$^{13}$ cm when they were on the
asymptotic giant branch (AGB).   The most plausible explanation  for the origin of these
disks is that an asteroid had its orbit sufficiently perturbed (Debes \& Sigurdsson 2002) that it entered the tidal zone of the white
dwarf and was shredded (Jura 2003).  Even with this scenario for disk formation,  we have very limited constraints on the disk masses, lifetimes, and compositions such as the
relative amounts of gas, rock and ice.   Here, we present new X-ray and infrared data which  enable an improved measure of the evolution of the disks orbiting two  accreting white dwarfs and thus enhance their use as probes of extrasolar planetary systems.

G29-38 and GD 362 were the first and second single white dwarfs, respectively, found to have an infrared excess (Zuckerman \& Becklin 1987, Becklin et al. 2005, Kilic et al. 2005).   As with other DA (hydrogen-dominated) white dwarfs, only a few heavy elements have been identified in the atmosphere of G29-38 (Zuckerman et al. 2003).  However, GD 362's atmosphere is helium-dominated and this star displays a rich spectrum (Gianninas et al. 2004) with  17 identified  elements (Zuckerman et al. 2007).  With silicon as a standard, GD 362 is enhanced in refractories such as calcium and titanium and deficient in volatiles such as sodium and carbon. 

 Reach et al. (2009) present two  models to account for G29-38's 5-40 ${\mu}$m infrared spectrum; their optically thin model requires water-ice while the optically thick model does not.  It is  possible, but unproven, that white
dwarf disks possess hydrogen as well as dust.
With X-ray observations of G29-38, we can assess
the amount of gas this hydrogen-dominated star accretes.  

The shape of its 10 ${\mu}$m silicate feature indicates that G29-38 is orbited by  dust from a disrupted comet or asteroid rather than from the
interstellar medium (Reach et al. 2005, 2009).  However, these data, by themselves, do not allow us to decide whether the dust originates near the star from a tidally-disrupted  asteroid (Jura 2003) or, instead,  arises in a cold outer reservoir and then drifts inwards to replenish the observed warm disk, as occurs around some main-sequence stars (Chen et al. 2006).  Here,
we  report
70 ${\mu}$m observations of G29-38 which  constrain the rate of inward drift of dust grains from an outer cold reservoir onto
the warm disk. 

Two  
recent studies of white dwarfs with helium-dominated photospheres propose that much of their atmospheric hydrogen arises from accretion of interstellar matter (Dufour et al. 2007, Voss et al. 2007).  GD 362 has an anomalously high hydrogen content, and we consider if this could have resulted from  accretion of an object or objects with a substantial amount of interior water.

In ${\S2}$ we describe  our X-ray observation show how these data can be used to constrain
the gas accretion rates.  In {\S3} we report our 70 ${\mu}$m data and describe how they can be used to constrain the rate of dust infall from a cold reservoir.    In $\S4$, we  describe observations and constraints on the hydrogen in GD 362's photosphere.  In $\S5$ we discuss our
results and in $\S6$ we present our conclusions.

\section{X-RAY RESULTS}
\subsection{Data}

Using {\it XMM-Newton}, G29-38  was observed for
23 ks starting on 2005 November 28 at 21:04:33 UTC, and GD 362
 was observed for 5 ks starting on
2006 September 4 at 02:06:23 UTC and for 14 ks starting on 2006
September 8 at 01:50:55 UTC.  We analyzed the data taken with the
European Photon Imaging Camera (EPIC) using standard
procedures.\footnote{{\tt
http://xmm.vilspa.esa.es/external/xmm\_user\_support/documentation/sas\_wsg/USG/USG.html}} We processed the observation
data files using the tools {\tt epchain} and {\tt emchain} from the
Science Analysis Software version 7.0. The events were filtered to
remove events near the edges of the detector chips and bad pixels, and
to reject events that were likely to be cosmic rays. We then examined the data for solar 
flares in the particle background, during which the detector counts from 
each array were 2 standard deviations above the mean count rate. Flares 
lasting $\approx$5 ks
were found in all detectors during the observation of G29-38, and one
lasting 1 ks was found in the 2006 September 4 observation of GD
362. We ignored the data taken during the flaring intervals, leaving
exposures of G29-38 totaling 17.1 ks in the pn detector (Struder et al. 2001) and 18.5 ks in each
MOS detector (Turner et al. 2001), and of GD 362 totaling 18.1 ks in the pn detector and 20.7 ks in each
MOS detector.

We  found that neither source 
was detected in any of the exposures. We 
computed upper limits to the net counts received by comparing the 
number of photons received within a 15\arcsec\ radius circle centered on the 
source to the number received within a 1\arcmin\ radius circle placed on blank 
sky 1.5\arcmin\ away from of each source, using the
algorithm described in Kraft et al. (1991). The counts in the on-source
region, the off-source region, and the derived 
net counts are listed in Table 1. 

G29--38 is located 
only 15\arcsec\ from an unidentified X-ray point source 
($\alpha, \delta$ = 23$^{\rm h}$ 28$^{\rm m}$ 46$.\!^{\rm s}$7, 
+05$^\circ$ 14\arcmin\ 50\arcsec) with 30 ${\pm}$ 10 counts in both the pn and the combined MOS detectors, which slightly complicated our 
derivation of an upper limit.  As can be seen in Table 1, there is marginal evidence for an excess in the pn detector, but no such evidence in the MOS detector.  However, the pn detector has much larger pixels than the MOS detector (4{\farcs}1
versus 1{\farcs}1 ), so we conclude that these excess
counts are an artifact of uncertainties in the shape of the Point Spread Function. 
We converted the count rates to fluxes using the Portable Multi-Mission 
Simulator,\footnote{{\tt http://heasarc.gsfc.nasa.gov/Tools/w3pimms.html}}
and list them in Table 1. 
For the 0.3--2.0 keV band, we assumed that the emission could be 
described by a 0.5 keV plasma. For the 0.3--10 keV band, we assumed
that it would originate from 5 keV thermal bremsstrahlung. 
Interstellar absorption is negligible toward both sources.  

\subsection{Implications}
Accretion onto white dwarfs can produce X-ray emission, and we use our observational upper limits  to constrain
the mass flux onto the observed stars.  Szkody et al. (2004) and Schmidt et al. (2007) have investigated  X-ray emission from white dwarfs with accretion rates as low as 4 ${\times}$ 10$^{12}$ g s$^{-1}$, and we follow
their analysis.  We thus assume that the accretion flows onto G29-38 and GD 362 are in the ``bombardment regime" described by
Kuijpers \& Pringle (1982).  In such an environment, a gas layer with a temperature near 10$^{7}$ K is formed, and this hot gas  freely radiates
away the accretion energy.  Consequently, because half of the emitted photons are radiated into the photosphere and other photons are emitted at wavelengths outside of the {\it XMM-Newton} MOS passband, we make
the very simple approximation for the X-ray luminosity, $L_{X}$, that (Patterson \& Raymond 1985) 
\begin{equation}
L_{X}\;{\approx}\;\frac{1}{4}\,\frac{G{\dot M}_{X}\,M_{*}}{R_{*}}
\end{equation}
Here, ${\dot M}_{X}$ is the mass accretion rate inferred from X-rays, and $M_{*}$ and $R_{*}$ are the mass and radius of the white dwarf, respectively.

We list in Table 2, our adopted values of the stellar masses, radii and distances (Kilic et al. 2008a, Reach et al. 2009, Farihi et al. 2009) as well as the inferred value of   ${\dot M}_{X}$ as derived from Equation (1) and the X-ray fluxes listed in Table 1.  Our X-ray data for G29-38 show that the upper limit to the  mass accretion rate of 2 ${\times}$ 10$^{9}$ g s$^{-1}$ is consistent with the heavy element accretion rate of 5 ${\times}$ 10$^{8}$ g s$^{-1}$ required to explain this star's
composition (Farihi et al. 2009) and inconsistent with the accretion rate of 5 ${\times}$ 10$^{10}$ g s$^{-1}$ expected for hydrogen-rich interstellar matter (Koester \& Wilken 2006).  Our result also is consistent with the estimate by Reach et al. (2009) that the mass in water-ice is 1/4 the mass in olivine in the material orbiting G29-38.    The   settling time for heavy elements in G29-38's hydrogen-dominated atmosphere  is less than one year  (Koester 2009), and therefore
a steady state model is appropriate.  However, the characteristic heavy element  settling time in GD 362's atmosphere is  ${\sim}$10$^{5}$ yr  (Table 3 and Koester 2009).  In {\S4} we present time-dependent models for the accretion that are consistent with the upper bound to the X-ray luminosity.  The X-ray limit is also consistent with the sum of the accretion rates derived from Equation (10) in ${\S4.2}$ and listed in Table 3.
\section{INFRARED RESULTS}
\subsection{Data}
 Previously, G29-38   has been observed only at ${\lambda}$ $<$ 40 ${\mu}$m (Reach et al. 2009).  Here we report observations of  G29-38 at 70 $\mu$m that were executed on 28 July 2008 with the 
Multiband Imaging Photometer for {\em Spitzer} (MIPS; Rieke et al. 2004, Werner et al. 2004).  
Using the default camera settings and small field of view, the white dwarf 
was imaged over 40 cycles of the MIPS dither using 10 s exposures and the default dither pattern, 
for a total integration time of 4.4 ks.

Owing to the expected target faintness, the data reduction was performed 
on the median-filtered basic calibrated data frames, which produced a higher
quality mosaic than the non-filtered frames, in this instance.  The mosaics were
created using MOPEX\footnote{http://ssc.spitzer.caltech.edu/postbcd/mopex.html},
and photometry was performed on the final image produced from the filtered frames.
There was no apparent source of flux at the expected location of the white dwarf,
and this was verified by aperture photometry executed with a radial 16$''$  (4 pixel)
aperture and a sky annulus between $18''$ and $39''$ from G29-38. The sky noise per pixel was measured
within the 4 pixel aperture radius, multiplied by the square root of the number
of aperture pixels, and multiplied by the average of the 60 K and 10 K aperture
corrections (2.185) found in the MIPS 
Data Handbook version 3.3.1 (Spitzer 
Science Center 2007).  The resulting $3\sigma$ upper limit is 1.32 mJy.

\subsection{Implications}

We  use the upper limit to the 70 ${\mu}$m flux from G29-38 to place an upper limit to the rate of dust infall from a cold reservoir, ${\dot M}_{dust}$.  If the particles orbit at some initial radius, $D_{init}$, and subsequently drift inwards because of Poynting-Robertson drag until they reach a final radius, $D_{final}$, then the flux, $F_{\nu}$ at frequency ${\nu}$ is (Jura et al. 2007a):
\begin{equation}
F_{\nu}\;{\approx}\;\left[\frac{1}{2}\,\ln\left(\frac{D_{init}}{D_{final}}\right)\frac{{\dot M}_{dust}\,c^{2}}{{\nu}}\right]/(4{\pi}\,D^{2})
\end{equation}
where $D$ is the distance from the Earth to the white dwarf.   We assume that the reservoir lies at 5 AU where the parent bodies might have a chance of surviving the star's AGB evolution (Jura 2008)  and that the grains drift inwards until they reach the outer boundary of the circumstellar disk  near 10$^{10}$ cm (Jura 2003, Reach et al. 2005, 2009).  Therefore,  we adopt $D_{init}/D_{final}$ ${\approx}$ 10$^{4}$, although  our results are insensitive to this ratio which only enters logarithmically into the estimate for
${\dot M}_{dust}$.  With $F_{\nu}$(70 ${\mu}$m) ${\leq}$ 1.3 mJy, we find that ${\dot M}_{dust}$ ${\leq}$   10$^{6}$ g s$^{-1}$, a factor of 500 less than the current heavy element accretion rate.    We conclude the disk is not being replenished under the action of Poyting-Robertson drag from
a reservoir of cold dust.  A plausible alternative is that the disk is derived from a tidally-disrupted asteroid
(Jura 2003).

\section{GD 362}

\subsection{Hydrogen to Helium Ratio}

Zuckerman et al. (2007) found that GD 362's atmosphere has an appreciable  mixture of  hydrogen and helium. Because accurate  models for the outer layers of white dwarfs with such a mixed composition were not available, they  could only report the abundance ratio of hydrogen to helium but not the total amount of hydrogen.   Subsequently, Koester (2009)  computed
a detailed model for GD 362's outer envelope, and, using the observations of Zuckerman et al. (2007), found that there is 7.0 ${\times}$ 10$^{24}$ g of 
the hydrogen in the star's mixing layer.  We now consider possible explanations.

The relative amount of hydrogen and helium in the atmospheres of helium-dominated white dwarfs is
 not fully understood (Tremblay \& Bergeron 2008, Chen \& Hansen 2009).  Regardless, it appears that GD 362 is unusual.
  Zuckerman et al. (2007) measured $n$(He)/$n$(H) = 14  in GD 362's photosphere, and there are two possible scenarios to explain
 this measurement. First, the star may have been a hydrogen-dominated white dwarf until it cooled to an effective temperature less than
 11000 K.  At this temperature, the mass of the convective envelope increased, and it is possible that  some interior helium was dredged up to the surface (Tremblay \& Bergeron 2008). However, even if this scenario correctly describes GD 362's
  history, the star still may be unusual.   Bergeron et al. (1990) analyzed
 37 DA white dwarfs and with the assumption that all the stars have $\log$ $g$ = 8.0, they found  that most have $n$(He)/$n$(H) $<$ 1, but deduced that  G1-7 and G67-23 have $n$(He)/$n$(H) ${\approx}$ 20, near the
 ratio determined for GD 362.  
However, these two abundance ratios are questionable.  Helium absorption lines in G1-7 and G67-23 are not directly measured; instead helium abundances are inferred from the hydrogen line profiles.  Broadening  caused by  an increased helium abundance and broadening  stemming from
 an increased surface gravity are degenerate (Zuckerman et al. 2003), and the assumed value of $\log$ $g$ for these two stars are probably too low.     For G1-7, Liebert et al. (2005) derived $\log$ $g$ = 8.83, while for G67-23,  Zuckerman et al. (2003) and Holberg et al. (2008a) derived  $\log$ $g$ = 8.5 and 8.8, respectively.  The trigonometric parallax for G67-23 supports the high gravity model (Holberg et al. 2008b).  Therefore, the photospheric
 value of $n$(He)/$n$(H) in GD 362 may be uniquely high for a DA white dwarf.

A second possibility is that GD 362 has been evolving continuously as a  star with a helium-dominated atmosphere.      In Figure 1, we display the hydrogen masses in the convective zones of helium-dominated white dwarfs from Voss et al. (2007) and Dufour et al. (2007) for stars in the temperature range 8000 K ${\leq}$ T ${\leq}$ 20,000 K.  Dufour et al. (2007) report values of $n$(H)/$n$(He), and we use these ratios along with  the total 
mass in the convective zone for helium-doiminated white dwarfs from Koester (2009), a value which  typically is near 2 ${\times}$ 10$^{28}$ g, to determine the mass of hydrogen in the white dwarf's atmosphere.   
 For reference, we also plot the amount of hydrogen in Ceres and Callisto by assuming  water contents of 25\% by mass and 50\% by mass, respectively (Michalak 2000, McCord \& Sotin 2005, Thomas et al. 2005, Canup \& Ward 2002), and, as standard in these models, by assuming that the hydrogen is largely contained within H$_{2}$O.
We see from Figure 1 that  the total mass of hydrogen in GD 362's convective envelope  appears anomalously large. GD 16 is another helium-dominated star with a distinctively large amount of hydrogen and an infrared excess.  Both Voss et al. (2007) and Dufour et al. (2007)
attribute much of the hydrogen in the atmospheres of white dwarfs with helium-dominated atmospheres to  accretion from the interstellar medium.  Because GD 362 has both an infrared excess and a substantial atmospheric pollution of heavy atoms, we consider models where its hydrogen is derived from a tidally-disrupted parent body.
During the star's red giant evolution before reaching the  AGB and becoming a white dwarf, it sublimated  surface water-ice from any orbiting parent body within at least 40 AU (Jura 2004).   
We imagine that  the parent body or bodies that have been accreted by GD 362 had
internal water to account for the atmospheric hydrogen as, for example,  in models of Callisto (Canup \& Ward 2002).

\subsection{ Schematic Model}

We  model  GD 362's atmospheric abundances of hydrogen and heavy elements, as well as the upper limit to the current accretion rate derived in ${\S2.2}$. 
  We assume that a parent body  of mass, $M_{par}$, has its orbit perturbed so that it is tidally
disrupted by the white dwarf (Debes \& Sigurdsson 2002, Jura 2003), and this defines the starting time, $t$ = 0, of the accretion event.    The  debris is rapidly reduced by mutual collisions to small dust particles, some at least as small as 1 ${\mu}$m in radius to account for the strong silicate emission (Jura et al. 2007b), and the material forms a disk. Material in this  disk then accretes onto the star with
a characteristic time scale $t_{disk}$ that is determined by viscous dissipation.  We write for the mass of the disk, $M_{disk}(t)$, that:
\begin{equation}
M_{disk}(t)\;=\;M_{par}\,e^{-t/t_{disk}}
\end{equation}
The accretion rate onto the star, ${\dot M_{*}}$, equals $-{\dot M}_{disk}$ or:
\begin{equation}
{\dot M_{*}}(t)\;=\;\frac{M_{par}}{t_{disk}}\,e^{-t/t_{disk}}
\end{equation}

Because hydrogen rises to the top, the total mass of hydrogen in  GD 362 is the amount in the mixing layer, $M_{mix}(H,t)$, and:
\begin{equation}
M_{mix}(H,t)\;=\;{\int}_{0}^{t}\,\,{\dot M}_{*}(H,t')\,dt'\;=\;M_{par}(H)\,\left(1\,-\,e^{-t/t_{disk}}\right)
\end{equation}
In ${\S4.3}$,  we use observational constraints to derive the composition of the parent body as a function of $t$ and $t_{disk}$.
Thus:
\begin{equation}
M_{par}(H)\;=\;\frac{M_{mix}(H,t)}{1\,-\,e^{-t/t_{disk}}}
\end{equation}

The total mass of heavy element $Z$ in the mixing layer of the star, $M_{mix}(Z,t)$, is governed by both the gain from accretion and the loss from settling which are described by Dupuis et al. (1993) as:
\begin{equation}
\frac{dM_{mix}(Z,t)}{dt}\;=\;{\dot M}_{*}(Z,t)\,-\,\frac{M_{mix}(Z,t)}{t_{set}(Z)}
\end{equation}
In Equation (7) $t_{set}(Z)$ denotes the time that element $Z$ takes to settle out of the surface mixing layer.  With the boundary condition that $M_{mix}(Z,t)$ = 0 at $t$ = 0, then:
\begin{equation}
M_{mix}(Z,t)\;=\;\frac{M_{par}(Z)\,t_{set}(Z)}{t_{disk}\,-\,t_{set}(Z)}\,\left(e^{-t/t_{disk}}\,-\,e^{-t/t_{set}(Z)}\right)
\end{equation}
We define a time-related parameter, ${\tau}(Z)$, which can be either positive or negative, for each element:
\begin{equation}
{\tau}(Z)\;=\;\frac{t_{disk}\,t_{set}(Z)}{t_{disk}\,-\,t_{set}(Z)}
\end{equation}
With Equations (4) and (9), Equation (8) becomes:
\begin{equation}
M_{mix}(Z,t)\;=\;{\dot M}_{*}(Z,t)\,{\tau}(Z)\,\left(1\,-\,e^{-t/{\tau}(Z)}\right)
\end{equation}
In ${\S4.3}$, we infer the composition of the parent body from  $M_{mix}(Z,t)$.  Thus, from Equations  (8) -- (10):
\begin{equation}
M_{par}(Z)\;=\;\frac{M_{mix}(Z,t)\,t_{disk}\,e^{t/t_{disk}}}{{\tau}(Z)\left(1\,-\,e^{-t/{\tau}(Z)}\right)}
\end{equation}
 To compute the  mass of oxygen in the parent body, we assume this element  is carried in water and
oxides.  Thus:
\begin{equation}
M_{par}(O)\;=\;\frac{m_{O}}{2\,m_{H}}\,M_{par}(H)\,+\,{\sum}\,\frac{n(O)\,m_{O}}{n(Z)\,m_{Z}}\,M_{par}(Z)
\end{equation}
where the mean atomic weight of each element is denoted by $m_{Z}$ and the $n's$ denote the number of atoms in the oxide
$Z_{n(Z)}O_{n(O)}$.  

To illustrate our  model, we consider various limiting cases which are seen in detail in ${\S4.3}$.   If $t$ $<<$ $t_{set}(Z)$ and $t$ $<<$ $t_{disk}$, then Equation (11) can be approximated:
\begin{equation}
M_{par}(Z)\;{\approx}\;M_{mix}(Z,t)\,\frac{t_{disk}}{t}
\end{equation}
Thus, at early times, a large parent body mass is required to reproduce the observed pollution because only a small fraction of its initial mass has been accreted.  In the case
with
$t$ $>$ $t_{disk}$ and $t_{disk}$ $>>$ $t_{set}$, then Equation (11) becomes:
\begin{equation}
M_{par}(Z)\;{\approx}\;M_{mix}(Z,t)\,\frac{t_{disk}}{t_{set}(Z)}\,e^{t/t_{disk}}
\end{equation}
At long times a large parent body mass is required because most its mass has accreted and settled below the mixing layer.
At intermediate times, $M_{par}(Z)$ achieves a minimum.  In the circumstance where $t_{disk}$ is only slightly larger than $t_{set}(Z)$, then 
as seen in Equation (9), ${\tau}(Z)$ becomes relatively large compared to $t_{set}$.  Thus, when $t$ $>>$ ${\tau}(Z)$,  by Equation (10), $M_{mix}(Z,t)$  is enhanced over cases where ${\tau}(Z)$ ${\approx}$ $t_{set}(Z)$.

In ${\S4.3}$, we compare the sum of the accretion rates for each element, ${\dot M}_{*}(Z,t)$, as  derived from the optical data, with the upper bound from the X-ray data. When $t$ $>>$ ${\tau}(Z)$ and ${\tau}$ $>$ 0, then from Equation (10),
\begin{equation}
{\dot M}_{*}(Z,t)\;{\approx}\frac{M_{mix}(Z,t)}{{\tau}(Z)}
\end{equation}
In the limit that $t_{disk}$ $>>$ $t_{set}$, Equation (15) becomes:
\begin{equation}
{\dot M}_{*}(Z,t)\;{\approx}\;\frac{M_{mix}(Z,t)}{t_{set}(Z)}
\end{equation}

\subsection{Application to GD 362}

If GD 362 has always had a helium-dominated atmosphere,   two sorts of models may explain its atmospheric hydrogen.
 One possibility is that the hydrogen
was accreted previous to the era when the currently orbiting disk was formed.   This earlier accretion could have been from either one large parent body or a swarm of asteroids.  Alternatively, one event may account for all of GD 362's distinctive properties, and using the formalism of 
${\S4.2}$, we now describe such a model.  

The observational constraints to be matched are the observed atmospheric abundances of
hydrogen and detected elements heavier than oxygen contributing at least 1\% of the accreted mass, and the upper limits to the mass accretion rate and the atmospheric oxygen abundance.  As in the Earth (Palme \& O'Neill 2004), we assume  that within the parent body, aluminum, magnesium, silicon and calcium are contained in the oxides MgO, Al$_{2}$O$_{3}$, SiO$_{2}$ and CaO while iron and nickel are  largely metallic.  Although oxygen is not detected in GD 362's photosphere, our expectation of a parent body composed substantially of oxides is consistent with the fit of GD 362's 10 ${\mu}$m  emission feature by grains with an olivine stoichiometry (Jura et al. 2007b).   As in the Earth's interior, we assume that hydrogen is largely contained within H$_{2}$O (Wood et al. 1996).  Therefore, the  7.0 ${\times}$ 10$^{24}$ g of hydrogen in GD 362's implies  the parent body must have
possessed at least 5.6 ${\times}$ 10$^{25}$ g of oxygen.  Table 3 provides  details.  

A major uncertainty is the disk lifetime, $t_{disk}$.  Jura (2008) proposed that $t_{disk}$ = 1.5 ${\times}$ 10$^{5}$ yr by estimating the viscosity in a  disk composed mostly of dust with some additional gas. Kilic et al. (2008b) suggested a value of $t_{disk}$ to be ${\sim}$10$^{5}$ yr since HS 2253+803, a heavily polluted white dwarf with a helium-dominated atmosphere, does not have an infrared excess.  Von Hippel et al. (2007) and Farihi et al. (2008) argued that disks composed purely of dust may survive over 10$^{9}$ yr.  All these estimates are very uncertain and $t_{disk}$ may vary from one disk to another.  We therefore consider models where $t_{disk}$ ranges from  2${\times}$ 10$^{5}$ yr to 10$^{8}$ yr.  We truncate all the calculations at $t$ $>$ 30 $t_{disk}$ because, by this time, the material has largely dissipated.  We do not display results from models with $t_{disk}$ as low as 10$^{5}$ yr since such models do not match the data because, according to Equation (9), ${\tau}(O)$ is negative and the models
behave qualitatively differently from the ones we show.
  
To match the observational constraints, we follow the schematic model in ${\S4.2}$.  For any given
 value of $t$ and $t_{disk}$,  we first compute  $M_{par}$ from Equations (6), (11) and (12).  We then compute current values of  ${\dot M}_{*}$ from Equation (4) and $M_{mix}(O)$ from Equation (8).   The results are displayed in Figures 2-4.

  Figure 2 shows  the predicted mass of  oxygen in the mixing layer.    Because silicon's 
  abundance is more uncertain than that of most other heavy elements and because it is the most important carrier of oxygen among the heavy elements, we explicitly show  models for both the nominal and low values of $M_{mix}(Si)$ given in Table 3.  In all models, the expected amount of oxygen  initially is much greater than the observed upper limit because the parent body must be  largely  water  to explain the observed ratio of hydrogen to heavy elements.  However, at later times, most of the heavy
 elements accreted from  the parent body have settled below the mixing layer, and $M_{mix}(O)$ asymptotically nears or falls below the observational upper limit. 
  
 Figure 3 shows the model predictions for the total accretion rate  as a function of time.  Initially, the models all predict a high accretion
 rate because, as illustrated by Equation (13), a massive parent body is required.  Ultimately, at long times,  in all  models, ${\dot M}_{*}$  approaches an asymptotic value defined by Equation (15) or (16), and as listed in Table 3, is less than the upper limit derived from the X-ray data.  
  
Figure 4 shows the inferred mass of the parent body for  different models.  We also display on each curve the times
  when the model first achieves agreement with the limits on the total accretion rate and the mass of oxygen in the mixing zone.  
  Agreement is achieved with the data only when the parent body mass is greater than that of both Callisto and Mars. The minimum mass of
  the parent body required to satisfy the data is insensitive to the disk lifetime for $t_{disk}$ $>$ 10$^{6}$ yr.

We now estimate  heavy element composition in the parent body of the material accreted onto GD 362.  Since only models with $t$ $>>$ $t_{disk}$ fit the data,  we derive $M_{par}(Z)$  from Equation (16) and the parameters listed in Table 3 with the additional assumption that oxygen  accretes at its measured upper limit.   We show in Table 4 the results and compare with values for the bulk Earth.
   The inferred 6.3 ${\times}$ 10$^{25}$ g of internal water in the model GD 362 parent body is much greater than the 2 -- 8 ${\times}$ 10$^{24}$ g of internal water in the Earth (Wood et al. 1996).

 \section{DISCUSSION}

How can we better understand the parent body or bodies accreted onto GD 362?  Since the mass of the main-sequence
progenitor to GD 362 was ${\sim}$3 M$_{\odot}$ (Kilic et al. 2008b), the thermal history of its planet-forming nebula and  resulting planetesimals and planets likely were different from that of the Solar Nebula.  This might account for the relatively high abundance of calcium in GD 362's polluted photosphere.  In any case our unified model for GD 362's pollution with the assumption that magnesium, aluminum, silicon and calcium are contained within oxides could be falsified with  better measurement of the star's elemental  abundances, particularly oxygen and silicon.  Also, relative to the Sun, the Earth
is deficient in volatiles like chlorine, sulfur and phosphorus by factors of ${\sim}$10 to ${\sim}$100 (Palme \& O'Neill 2004).    Measurements of the abundances of these volatile elements could help clarify the
origin and evolution of the parent body responsible for polluting GD 362.  

It is possible that the atmospheric hydrogen in GD 362 was derived from the accretion of many small asteroids rather than one single large object.  The indicated minimum mass of the accreted material onto GD 362 of  ${\sim}$10$^{26}$ g  is much larger than the total mass of 2 ${\times}$ 10$^{24}$ g  of the Solar System's asteroid belt (Binzel et al. 2000).  However, an asteroid belt orbiting the A3V star ${\zeta}$ Lep  could be  as massive as 4 ${\times}$ 10$^{26}$ g (Chen \& Jura 2001) and such a system could supply the required mass.   
 
There exists evidence for the  destruction of massive rocky bodies  in extrasolar planetary systems.   
The main-sequence stars BD +20 307 and HD 23514 have  circumstellar dust (Song et al. 2005, Rhee et al. 2008) that likely resulted from 
collisions of  parent bodies of ${\sim}$10$^{26}$ g   (Rhee et al. 2008).    Ashwell et al. (2005) have proposed that the main-sequence F-type star J37 in the open cluster NGC 6633   has accreted  approximately 10$^{26}$ g of material from a disrupted planet.  A plausible model for the formation of the Moon is that a Mars-mass planetesimal collided with Earth when the Solar System had
 an age of ${\sim}$ 25 -- 30 Myr (Canup 2004).   
 
 After a planetary system's orbits have been drastically re-arranged during the AGB phase  mass loss, plausibly,  there could be a late perturbation of a Mars-mass planet.
 If the disk persists for  ${\sim}$10$^{8}$ yr, even very rare events might be salient.
 The scenario that the disk orbiting GD 362 is the result of the tidal-destruction of a Mars-mass planet is quite different from the suggestion by Reach et al. (2009) that the disk orbiting G29-38 might be debris from the core of a Jovian-mass planet that was engulfed during the AGB phase of the star's evolution.  The amount of hydrogen in
  GD 362's mixing layer  is a factor of ${\sim}$10$^{5}$ less than the mass of Jupiter.  Therefore there must have been very efficient
 but fine-tuned removal of  the bulk of a Jovian planet while retaining the rocky core if such a model is to explain GD 362's pollution.  

 We have raised the possibility that GD 362 has accreted from  a parent body with a mass at least as large as 10$^{26}$ g.  If a tidally-disurpted disk
was formed and  the dissipation time  was as short as  10$^{5}$ yr (Jura 2008, Kilic et al. 2008b), then the accretion rate
of 3 ${\times}$ 10$^{11}$ g s$^{-1}$  produced an X-ray luminosity of ${\sim}$ 5 ${\times}$ 10$^{27}$ erg s$^{-1}$.  Single white
dwarfs typically have X-ray luminosities less than 10$^{28}$ erg s$^{-1}$ (O'Dwyer et al. 2003), but there might be rare
exceptions that could be diagnostic of  accretion of a large parent body.

\section{CONCLUSIONS}

Our new X-ray and infrared data are consistent with the scenario that GD 362 and G29-38 are polluted  by
accretion of material from their circumstellar disks.  GD 362 also has an anomalously large mass of hydrogen. One possibility is that  before the current disk was formed, GD 362 was bombarded either by a swarm of Ceres-mass asteroids or by a single large object. An alternative  model to explain simultaneously
all of this star's distinctive properties is that we are witnessing the consequences of the tidal-destruction of a single parent body
at least as massive as Callisto and probably as massive as Mars.

This work has been partly supported by NASA and the NSF.  We thank the referee, F. Mullally, for a careful reading of the manuscript
and for asking helpful questions.

\newpage
\begin{center}
Table 1 -- X-Ray Counts 
\end{center}
\begin{center}
\begin{tabular}{lcccccccc}
\hline
\hline
Star  & Band & pn & pn & pn  & MOS&MOS&MOS& Flux \\
  & (kev) & On & Off &Net &On &Off &Net &  10$^{-15}$ erg cm$^{-2}$ s$^{-1}$ \\
  \hline
G29-38 & 0.3--2.0 & 17 & 9.3 & $<$15 & 7 & 6.3 & $<$6 & $<$1 \\ 
      & 0.3--10.0 & 23 & 15.2 & $<$16 & 12 & 11.5 & $<$7 & $<$2 \\ 
GD362  & 0.3--2.0 & 14 & 17.0 & $<$6 & 8 & 10.3 & $<$5 & $<$0.7 \\
      & 0.3--10.0 & 31 & 33.2 & $<$9 & 19 & 21.3 & $<$7 & $<$2 \\
      \hline
      \end{tabular}
      \end{center}  
Upper limits are 90\% confidence.  
\newpage
\begin{center}
Table 2 -- Stellar Properties
\end{center}
\begin{center}
\begin{tabular}{lccccccccc}
\hline
\hline
Star & $D$ & $M_{*}$ & $R_{*}$ &   ${\dot M}_{X}^{a}$ & $t_{cool}^{b}$\\
  & (pc) & (M$_{\odot}$) & (10$^{8}$ cm) &  (10$^{9}$ g s$^{-1}$) & (Gyr) \\
  \hline
G29--38 & 14 & 0.61 & 8.5 &   $<$2& 0.4 \\ 
GD 362  & 50 & 0.71 & 7.4 &   $<$20 & 0.7-0.8\\
\hline
\end{tabular}
\end{center}
$^{a}$see Equation (1)
$^{b}$white dwarf cooling age
\newpage
\begin{center}
Table 3 -- Heavy Elements in  GD 362's Convective Envelope
\end{center}
\begin{center}
\begin{tabular}{llllllllll}
\hline
\hline
Element & $M_{mix}(Z)^{a}$  & $t_{set}^{b}$ & ${\dot M}_{*}(Z,t)^{c}$ \\
 &   (10$^{21}$ g) &  (10$^{5}$ yr) & (10$^{9}$ g s$^{-1}$) \\
 \hline
 O & $<$11&1.13 & $<$3.1\\
 \\
  Mg& 2.4$^{+1.9}_{-1.0}$ & 0.944& 0.81$^{+0.64}_{-0.34}$ \\
  \\
  Al &1.0$^{+0.65}_{-0.34}$  & 0.857 & 0.37$^{+0.24}_{-0.12}$ \\
  \\
Si  & 3.9$^{+3.9}_{-1.9}$&  0.795 & 1.6$^{+1.5}_{-0.80}$\\
\\
Ca  & 2.2$^{+0.60}_{-0.40}$ &  0.607& 1.2$^{+0.26}_{-0.26}$\\
\\
Fe  & 12$^{+3.0}_{-2.4}$&  0.467 &  8.2$^{+2.0}_{-1.7}$\\
\\
Ni &  0.48$^{+0.20}_{-0.14}$ & 0.445 & 0.34$^{+0.15}_{-0.10}$ \\
\\
Total$^{d}$ & 33.0 &  & 15.6$^{e}$ \\
\hline
\end{tabular}
\end{center}
$^{a}$current value from the relative abundances in Zuckerman et al. (2007) and the assumption that $M_{mix}(H)$ = 7.0 ${\times}$ 10$^{24}$ g (Koester 2009)  
$^{b}$from Koester (2009)
$^{c}$evaluated from Equation (16)
$^{d}$Assuming that the true oxygen value equals the upper limit
$^{e}$consistent with  the upper limit derived from the X-ray data of 2 ${\times}$ 10$^{10}$ g s$^{-1}$
\newpage
\begin{center} 
Table 4 -- Element  Fractional Composition by Mass in GD 362's Polluting Parent Body 
\end{center}
\begin{center}
\begin{tabular}{lllll}
\hline
\hline
Element & Parent Body$^{a}$ & Bulk Earth$^{b}$ \\
\hline
O &  0.20  & 0.32    \\
Mg &  0.052& 0.16  \\
Al & 0.024& 0.015\\
Si & 0.10& 0.17 \\
Ca &0.078 & 0.016\\
Fe & 0.53&0.28 \\
Ni &  0.022 & 0.016\\
\hline
\end{tabular}
\end{center}
$^{a}$from the accretion rates in Table 3 and ignoring minor constituents
$^{b}$from Allegre et al. (1995). The total fraction is less than 1.00 because we omit minor constituents.  
\newpage
\begin{figure}
\plotone{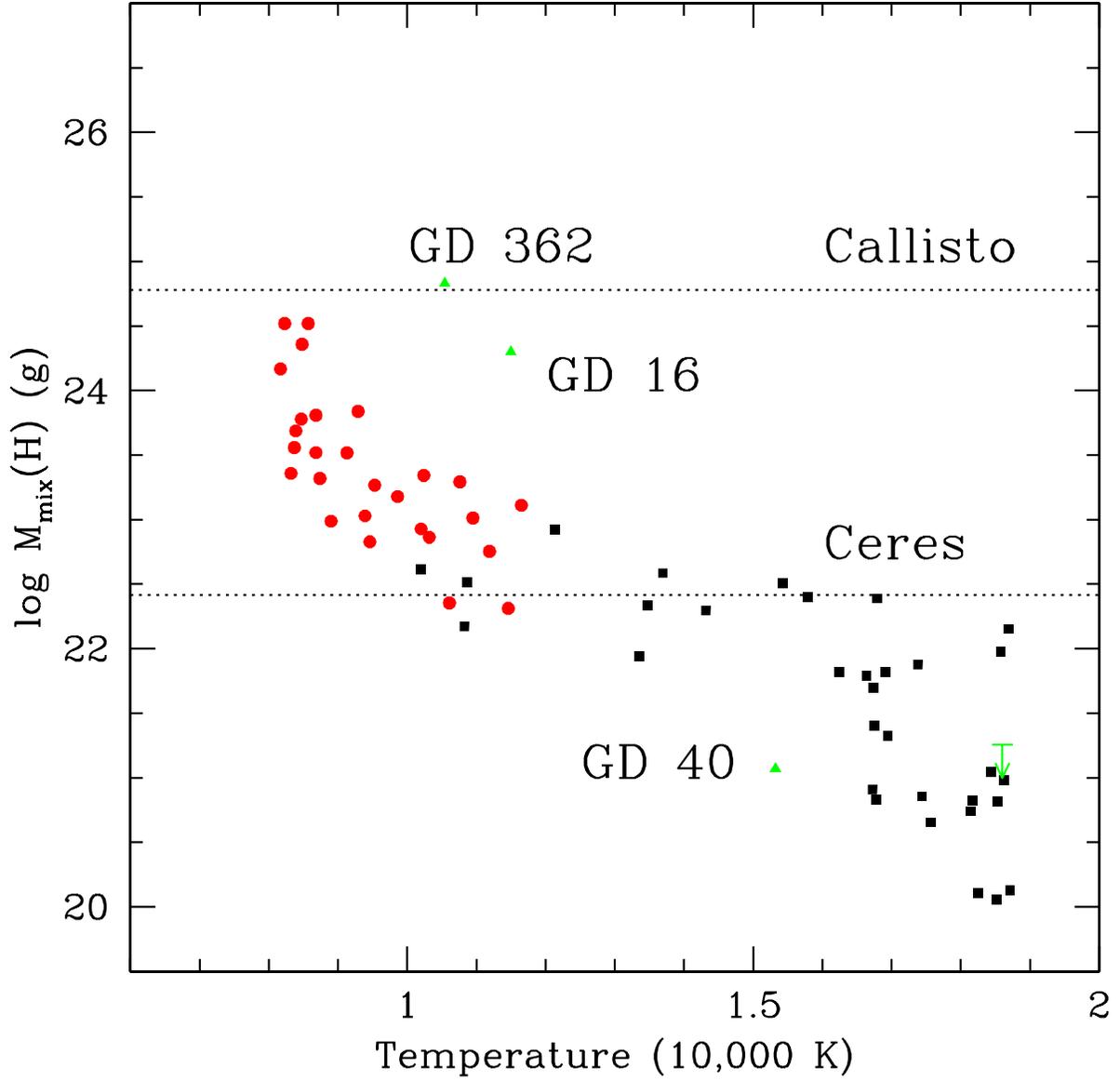}
\caption{Mixing envelope mass of hydrogen ($M_{mix}(H)$) for helium-dominated white dwarfs. Black squares are 
from Voss et al. (2007), red circles are from Dufour et al. (2007), and green triangles are white dwarfs with an infrared excess. The green upper limit refers to Ton 345 which is a heavy-element polluted helium-dominated white dwarf without detected hydrogen (Gaensicke et al. 2008) and with
an infrared excess (Melis et al. 2009). The atmospheric mass of hydrogen for GD 16 is taken from the observed hydrogen abundance (Koester et al. 2005)  and the calculation by Koester (2009, private communication) that the total mass in the star's convective envelope is 6.3 ${\times}$ 10$^{27}$ g.   The hydrogen masses for GD 362 and GD 40 are from Koester (2009) and Voss et al. (2007), respectively. The dotted lines show the estimated hydrogen masses in Callisto and Ceres (see $\S5$). }
\end{figure} 
\newpage
\begin{figure}
\plotone{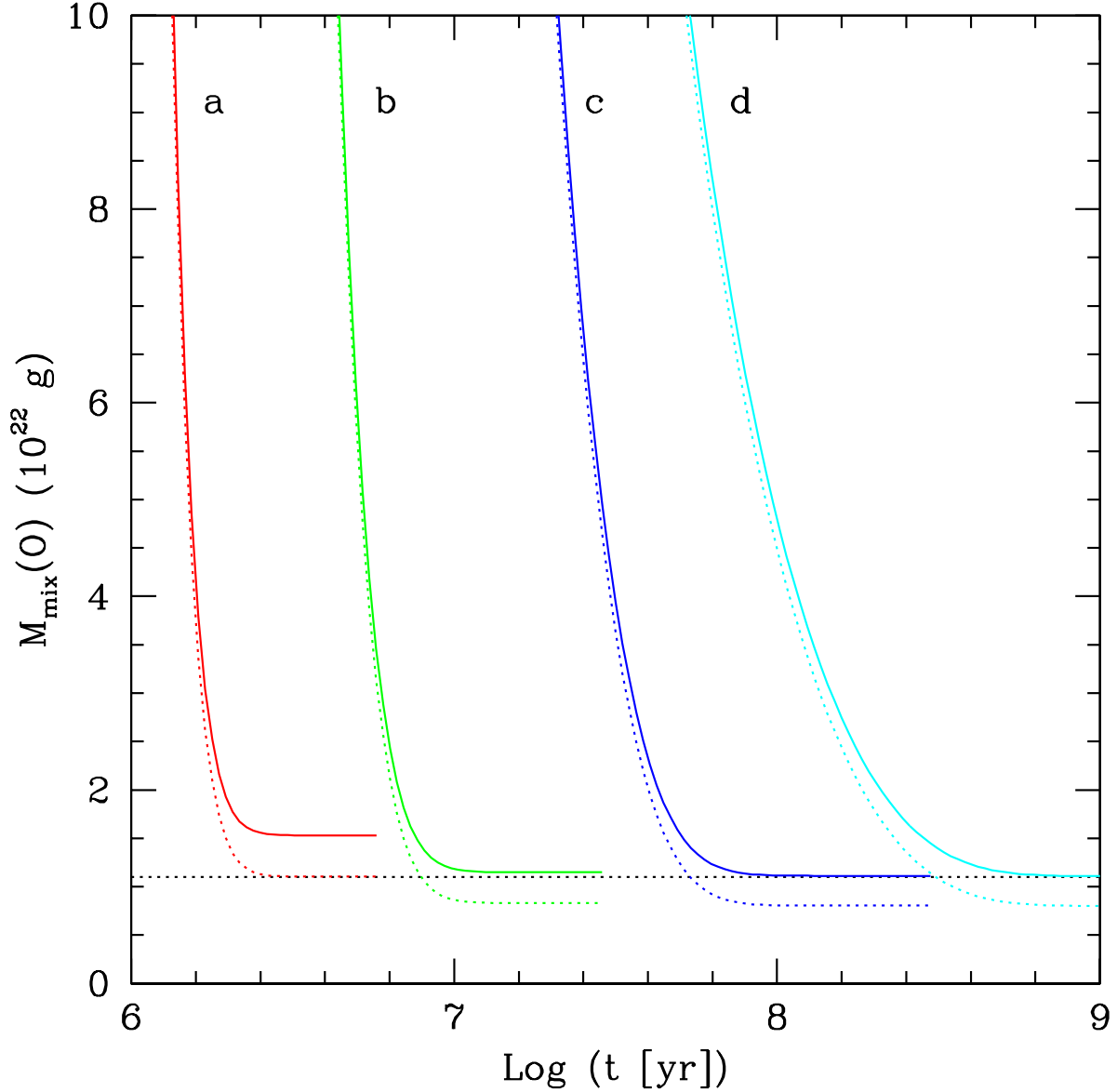}
\caption{Predicted mixing envelope mass of oxygen ($M_{mix}(O)$) in GD 362 as a function of time.  The red (model a), green (model b), blue (model c) and cyan (model d)
lines refer to values of $t_{disk}$ of 2 ${\times}$ 10$^{5}$, 10$^{6}$, 10$^{7}$  and 10$^{8}$ yr, respectively.  The solid and dotted lines refer to models where the silicon mass in the mixing zone is the nominal value and the lower bound, respectively, as listed in Table 3.  The dotted  horizontal black line  denotes the observational upper limit from Table 3.}
\end{figure}
\newpage
\begin{figure}
\plotone{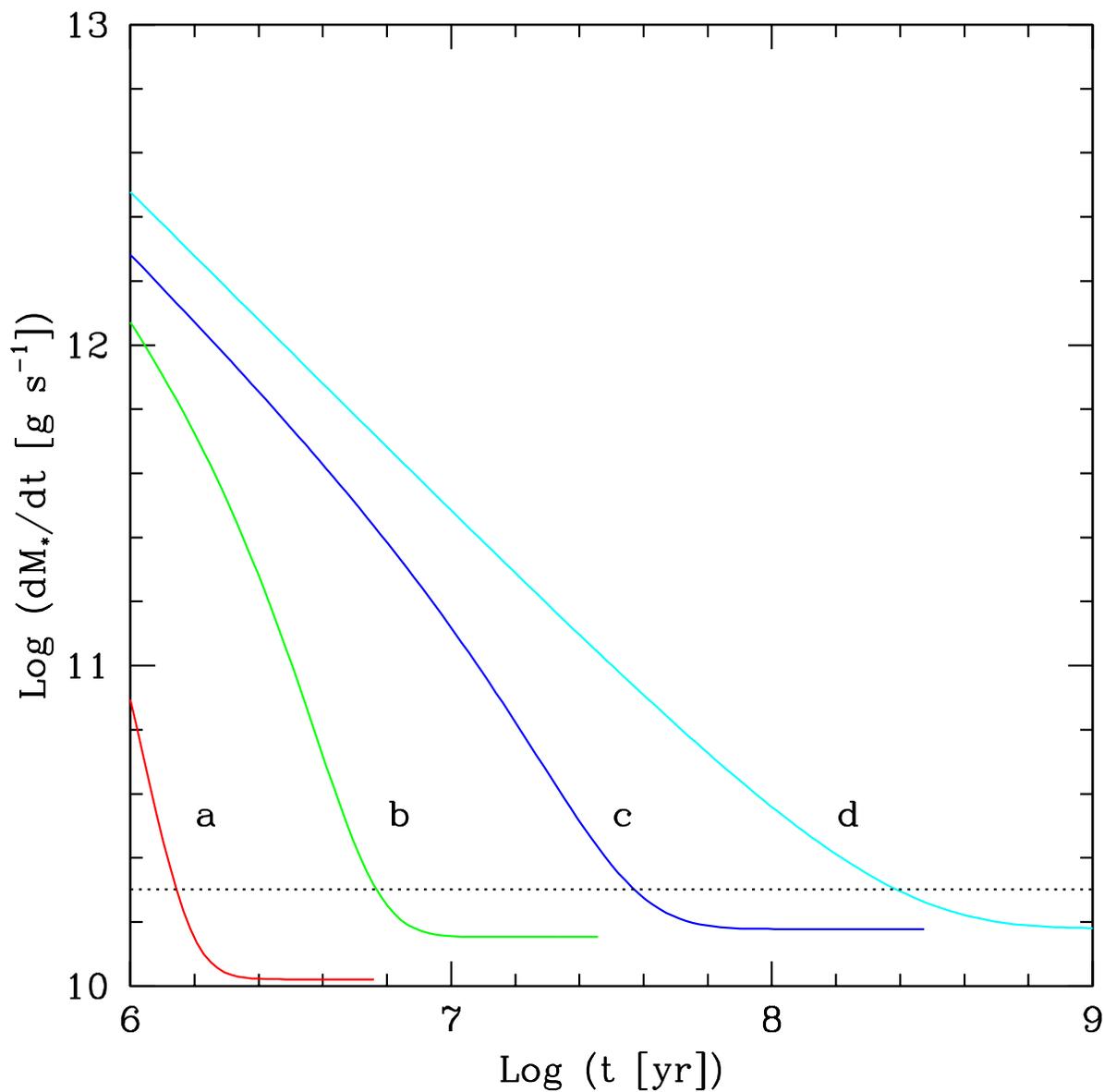}
\caption{Inferred mass accretion rates (${\dot M}_{*}$).  The different colored lines are defined  as in Figure 2.  Only solid curves are shown because the difference in ${\dot M}_{*}$ between the models with the nominal and low
silicon mass in the mixing zone are not significant.   The dashed horizontal line denotes the upper limit from our X-ray data.}
\end{figure}
\newpage
\begin{figure}
\plotone{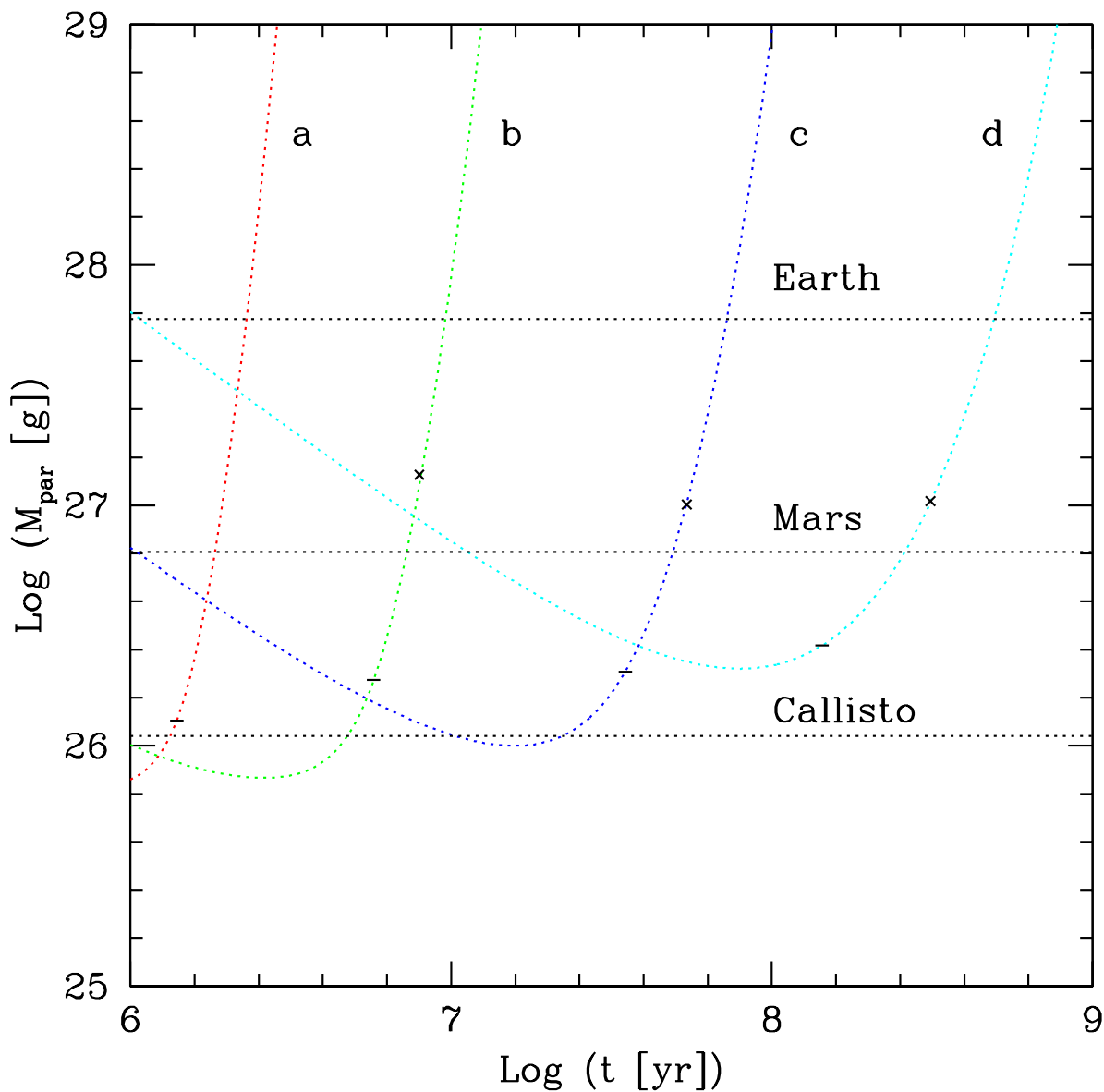}
\caption{Inferred total mass of the parent body ($M_{par}$) required to fit the observational constraints.
Each  colored line as defined  as in Figure 2 is marked by a solid black dash at the first  time when
the accretion rate falls below the inferred upper limit of 2 ${\times}$ 10$^{10}$ g s$^{-1}$ (as shown in Figure 3) and a black cross at the first time when the oxgyen
mass in the mixing zone falls below its upper limit of 1.1 ${\times}$ 10$^{22}$ g (as shown in Figure 2).   These results are displayed for the case using the lower
bound of the silicon mass in the mixing zone, the dotted curves in Figure 2.}
\end{figure}

\end{document}